\documentclass[twoside,twocolumn]{article}

\usepackage{blindtext} 

\usepackage[sc]{mathpazo} 
\usepackage[T1]{fontenc} 
\usepackage{microtype} 

\usepackage[english]{babel} 

\usepackage[hmarginratio=1:1,top=32mm,columnsep=20pt]{geometry} 
\usepackage[hang, small,labelfont=bf,up,textfont=it,up]{caption} 
\usepackage{booktabs} 

\usepackage{lettrine} 

\usepackage{enumitem} 
\setlist[itemize]{noitemsep} 

\usepackage[utf8]{inputenc}

\usepackage{abstract} 

\usepackage{titlesec} 
\renewcommand\thesection{\Roman{section}} 
\renewcommand\thesubsection{\roman{subsection}} 
\titleformat{\section}[block]{\large\scshape\centering}{\thesection.}{1em}{} 
\titleformat{\subsection}[block]{\large}{\thesubsection.}{1em}{} 

\usepackage{fancyhdr} 
\usepackage{titling} 

\usepackage{hyperref} 
\usepackage[square,sort,comma,numbers]{natbib} 
\usepackage{siunitx}
\usepackage{wrapfig}
\usepackage{graphicx}
\usepackage{framed}
\hypersetup{
    colorlinks=true, 
    linkcolor=blue,
    filecolor=magenta,      
    urlcolor=cyan,
}
\usepackage{url}

\usepackage{titling}
\usepackage{lmodern}
\usepackage{marvosym}

\usepackage[flushmargin]{footmisc}
\usepackage{setspace}

\begin{document}

\setlength{\droptitle}{-4\baselineskip} 
\pretitle{\begin{center}\huge\bfseries} 
\posttitle{\end{center}} 
\title{Cone-beam CT for Dental Radiography} 
\author{%
\textsc{Kaspar Höschel}\thanks{
TU Wien, Karlsplatz 13, 1040 Vienna, Austria
\newline\Letter ~ \href{mailto:kaspar.hoeschel@student.tuwien.ac.at}{kaspar.hoeschel@student.tuwien.ac.at}
}
%
%
%
\textsc{ \href{https://orcid.org/0000-0001-9881-7892
}{\includegraphics[scale=1.0]{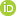}}}
\and
\textsc{Janos Juhasz}\thanks{Course instructor, PhD, MCCPM
\newline Medical Physicist
\newline Juravinski Cancer Centre at Hamilton Health Sciences, Hamilton, ON, L8V 5C2, Canada
} \\[1ex] 
}
\thanksmarkseries{arabic}
\date{}
\renewcommand{\maketitlehookd}{
\begin{abstract}
The cone beam computed tomography (CBCT) technique was first inserted in dental imaging 15 years ago. Due to this technique 3D imaging in dentistry has been excessively changed. The purpose beyond using CBCT instead of computed tomography (CT) imaging is that CT can not be used in many cases in dentistry,  
in addition to the higher radiation and cost of CT. Nowadays, CBCT can be applied in diagnostics for several dental specialties because of the variety of services it offers and the specification of use. In contrast to CT, CBCT has shown great benefits in endodontics, implant planning oral and maxillofacial surgery and even in orthodontics.
\end{abstract}
\setlength\parindent{15pt}\textbf{Keywords }Cone-beam computed tomography, Computed tomography, X-ray, Dental radiography
}
\maketitle 



\section{Introduction} 
\lettrine[nindent=0em,lines=3]{I}n dental radiography, the imaging technique CT
is mostly used to provide thin-slice images in the axial plane. CBCT makes it possible to acquire images in multiplanar reformation, which is not possible with CT. Therefore, obtaining images in the axial, coronal and oblique as well as sagittal planes with the practitioner has became a fact
\cite{scarfe2006clinical}.
\textit{Wilhelm C. Röntgen} modified \textit{Crookes tube} and discovered for the first time x-rays in 1895 \cite{crookes1879v,sansare2011early}.
The x-ray source transmits x-rays through the observed patient and the attenuated intensity is detected and converted into grayscales to obtain a dental or panoramic radiograph. 


In CT the x-ray source is rotating \ang{360} in a fan shaped beam around the patient. This fan beam cuts through the subject under study and the obtained tomograph shows 2D images which are based on reconstruction. In 1996, the first CBCT device (QR NewTom 9000) was introduced to the European Market by \textit{Tacconi et al.} \cite{gargcbct}. The CBCT uses a cone beam shaped x-ray bundle and the reconstruction technique to get a 3D image which allows the practitioner to find any plane or axis of the imaged anatomical structure. This sophisticated technique is applied in dental radiography.

The purpose of this review is to give an overview about CBCT in dental radiography. Starting with the physical background of CT and CBCT and tackling the techniques used as well as the  specific applications of CBCT in dental radiography will be discussed. 
\section{Physical background} 
\subsection{Image acquisition}
\begin{figure}[h]
    \centering
    \includegraphics[width=0.5\textwidth]{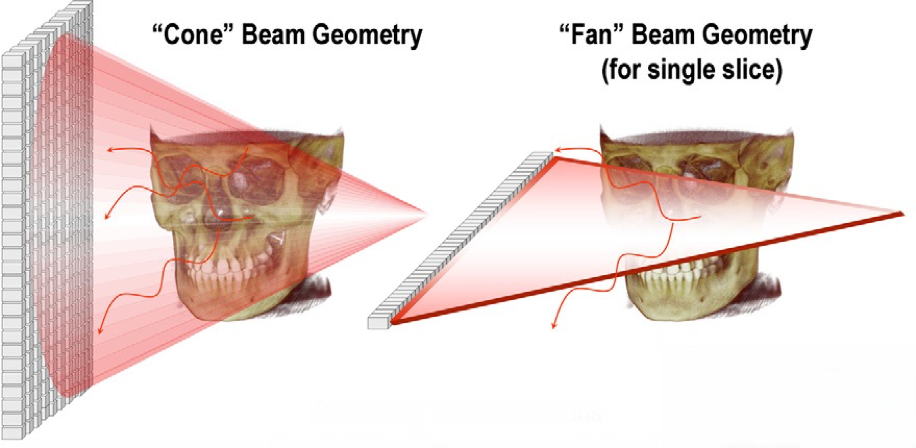}
    \caption{Beam geometry of CBCT (left) and CT (right) \cite{scarfe2008cone}}
    \label{fig1}
\end{figure}

The main difference between CBCT and CT depends on the beam geometry and its reconstruction (Fig. \ref{fig1}).

CT uses a narrow fan shaped beam and CBCT a cone beam shaped x-ray to irradiate the anatomic region which is under examination. In most CBCT systems the attenuated x-ray is detected by flatpanel detectors (FPD) which consist of a Caesium-Iodide scintillator (CsI) and converts the x-rays into visible light. Below the CsI are photosensitive pixels or diodes to absorb visible light. This light is converted into electric charge and transported by transistors (TFTs) to a read-out circuitry of the hardware or software. 

\subsection{Axial reconstruction}
The applied technique to reconstruct a 3D image in CBCT is called axial reconstruction. By stacking all images together and  the practitioner finds any plane and axis of the imaged anatomy (multiplanar reconstruction (MPR)) which is important in diagnosis.

The reconstruction technique rests upon the (FBP) algorithm (FBP) algorithm or the Feldkamp-Davis-Kress (FDK) algorithm to reconstruct 3D images of the obtained slices of the patient
\cite{li2006fdk,scarfe2006clinical}.

More precisely, this algorithm is based on the \textit{inverse Radon transform} to reconstruct a 3D image at any grayscale point along the x-ray beam. This principle includes filters to reduce e.g. smearing and blurring of the backprojected image. 

A CBCT image can be seen as a 3D volumetric dataset composed of voxels. Each voxel contains a certain attenuation coefficient. This dataset of voxels can be transversed into any plane of any axis \cite{semper2016}.

\section{Clinical applications and techniques} 

\subsection{Diplay mode techniques}
CBCT is effectively performed in maxillofacial imaging of hard and soft tissue. 
The used software allows the following  of various real-time display techniques from the volumetric data set \cite{scarfe2006clinical}:\\

\textit{\textbf{Oblique planar reformation}}\\ 
2D images are created at any angle when a set or stack of axial images are transected. This function allows to evaluate specific structures (e.g. impacted teeth or temporomandibular joint (TMJ) and is called nonorthogonal slicing images at any angle to obtain 2D planar images.\\

\textit{\textbf{Curved planar reformation}}\\
In this mode the long axis of the imaging plane is aligned with a particular anatomic structure. Hence, it shows a display of the dental arch and a panorama like view of thin-slice images.\\

\textit{\textbf{Serial transplanar reformation}}\\
This mode accomplished a set of sequential and stacked cross-sectional images perpendicular to the curved planar reformation. Thickness and spacing can be chosen individually (e.g. 1mm thick and apart).
The display mode is applied in the assessment of morphologic characteristics especially in alveolar bone height and width in implantology. Likewise, these images are used to evaluate the inferior alveolar canal related to impacted teeth or the characteristic of the symptomatic TMJ.\\

\textit{\textbf{Multiplanar volume reformations}}\\
This mode adds the adjacent voxels to provide an increased thickness of a multiplanar or "ray sum" image and relate precisely to the volume of the patient. Accordingly the slice thickness of curved planar reformatted images is increased to get panoramic or cephalometric images. Different from traditional radiographs these achieved images do not depict any distortions or magnification.\\

\textit{\textbf{Volume rendering}}\\
There are two rendering techniques: direct and indirect volume rendering. 
Direct volume rendering uses an arbitrary threshold below or above which precludes all gray values. Generally, an image is generated by the technique maximum intensity projection (MIP) is applied which takes only high density voxel values of a specific thickness above the threshold and excludes lower values. These images facilitate e.g. the location of impacted teeth, the assessment of fractures or TMJ. Further, applications are the analysis of craniofacial imaging or to visualize soft tissue clacifications.\\

In contrast, indirect volume rendering (IVR) selects the density of the voxels which are displayed in the total data set and results them in volumetric surface reconstruction with depth. This technique has two types of views: surface and volumetric rendering. Consequently, IVR can be applied to visualize and evaluate craniofacial conditions. Also, the inferior alveolar canal in connection with the mandibular third molar is analyzed 
\cite{venkatesh2017cone}.

\subsection{Specific applications in dental radiography}

\begin{figure}[h]
    \centering
    \includegraphics[width=0.45\textwidth]{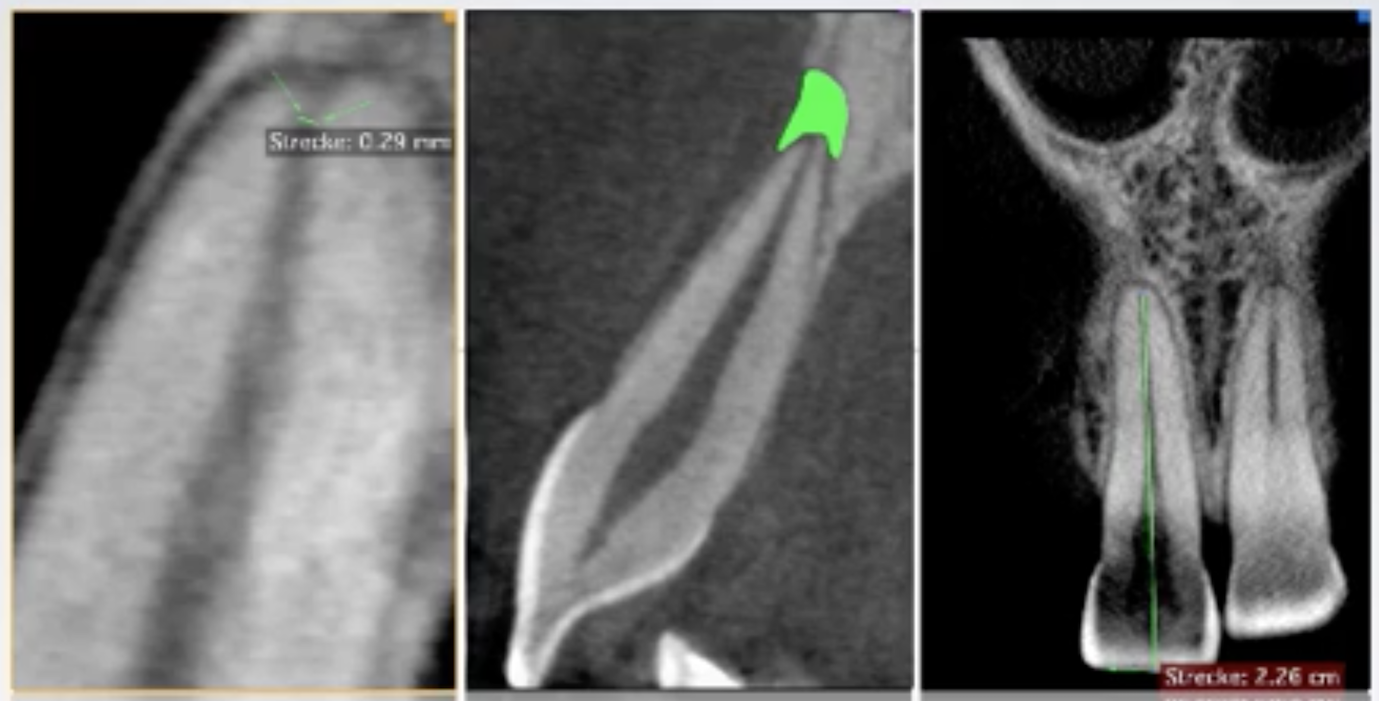}
    \caption{Endodontology \cite{semper2016}}
    \label{fig4}
\end{figure}

CBCT has shown great benefits due to performing length measurements in \textbf{endodontology}. Periapical lesions or the size of a foramen can be detected easily by volumetric estimation (Fig. \ref{fig4}) \cite{miles2013atlas,semper2016}.\\
More applications for CBCT scans in endodontics which are without magnification or distortion include:
\begin{itemize}
\item analysis of number, location of roots and connected canals,
\item type and degree of the root angle or the detection of vertical and horizontal root fractures \cite{venkatesh2017cone}.
\end{itemize}



\textbf{Orthodontics} focuses on correcting or avoiding the poor alignment or positioning of teeth, jaws, and face structure \cite{Fence2013}. Assessment of tooth position and eruption patterns from impacted teeth is complicated but can be facilitated with CBCT diagnostic and planning software (e.g. \textit{Dolphin 3D} or \textit{In Vivo Dental}). The software assists in the construction of a 3D virtual model and the conversion into standard medicine file format standard also known as '\textit{Digital Imaging and Communications in Medicine}' (DICOM) \cite{scarfe2008cone}.

Specific softwares for cephalometry in orthodontics allow facial growth analysis, dental age, airway functions or disturbances in tooth eruptions
\cite{miles2013atlas,semper2016,venkatesh2017cone}.\\


Another most common use of CBCT is \textbf{implantology} which shows remarkably evidences but it could be risky to set an implant without CBCT. In fact, if a doctor is using multiple implants for example in overdenture the preoperative implant site assessment is essential and allows precision in measurements of length, width and the angle of the implant \cite{miles2013atlas}. Axial and panaoramic views facilitate the implant site assessment \cite{venkatesh2017cone}.\\
In clinics radiographic stents are used with a nonmetallic marker for implant site location for many years. These radiographic stents can detect and evaluate with a precise location the bone receptor site. The clinician is analyzing the object under study while this nonmetallic marker helps in the evaluation \cite{carter2008american}.
Further, specific  diagnostic and planning software is used to enable virtual implant placements and build stents. 
Some companies allow computer assisted designs and produce implant prosthetics (e.g. NobelGuide/Pocera software)
\cite{semper2016}.\\


\textbf{Periodontolgy} describes the field of study of hard and soft tissues which provide support for teeth and preserve their jaw position
\cite{EFP2018}.
CBCT has not been researched adequately in periodontology yet but might be feasible in volumetric estimation of periodontal defects.
Misch et al. demonstrated that CBCT is an accurate measurement in analyzing periodontal diseases and intrabony defects. Also CBCT is more suitable in visualizing buccal and lingual defects than conventional 2D radiographs
\cite{misch2006accuracy,venkatesh2017cone}.

Another essential application for CBCT in dentistry is \textbf{oral and maxillofacial surgery}. CBCT allows accurate measuring of surface distances due to structural superpositions of 2D images. Consequently it alleviates the examination of particular fractures (e.g. midfacial and orbital fractures). Dentoalveolar fractures cannot be visible on 2D imaging but with CBCT scans
\cite{cevidanes2005superimposition,venkatesh2017cone}.

\begin{figure}[h]
    \centering
    \includegraphics[width=0.45\textwidth]{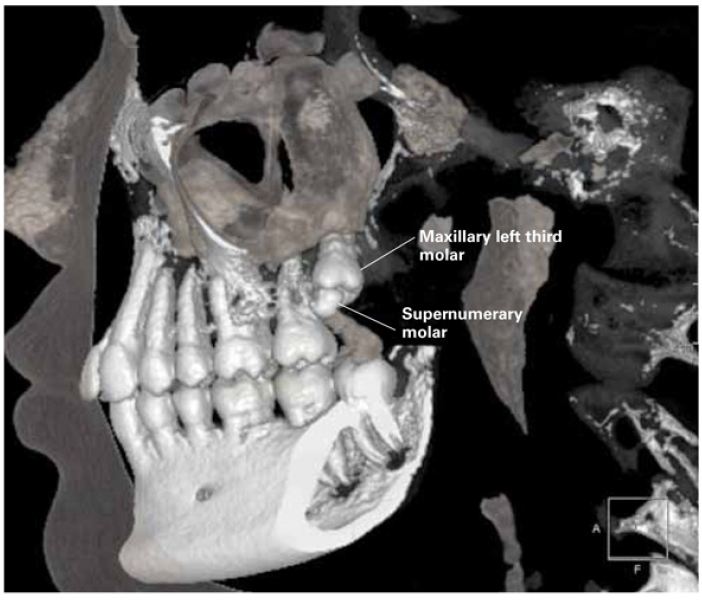}
    \caption{3D grayscale reconstruction \cite{miles2013atlas}}
    \label{fig5}
\end{figure}

Equally important is the \textbf{odontogenic lesion visualization} by using CBCT. Indeed, CBCT can detect lesions of odontogenic cysts and tumors. Furthermore, extension and position of pathologies and osteomyelitis can be determined. Fig. \ref{fig5} shows the 3D grayscale recontruction with CBCT used in odontogenics to illustrate both the maxillary left third molars and impacted or supernumerary teeth \cite{miles2013atlas,venkatesh2017cone}.\\

CBCT is applied in orthognathic or corrective jaw surgery. The outcomes for \textbf{orthognathic surgery}, \textbf{trauma evaluation} and \textbf{imaging} are 2D and 3D grayscale or even color information which have a high accuracy to identify anatomical structures such as soft tissue.\\


CBCT is effectively applied to characterize condylar changes and the appearance of the \textbf{TMJ complex} and shows more benefits in diagnosis and therapy. Conventional 2D gray scale techniques were used to image the TMJ complex. This technique is inappropriate to assess the variations in shape and size of the condyle from one side to another. With CBCT the images of condyle are multiplanar and in 3D shape and color. The TMJ and its function can be analyzed easily. Additionally, CBCT provides diagnosis of the glenoid cavity and location of surrounding soft tissue of the TMJ. Magnetic Resonance Imaging can be avoided or even neglected in applications of TMJ disorders by using CBCT
\cite{miles2013atlas,semper2016,venkatesh2017cone}.
%
\begin{framed}
Further applications of CBCT in dentistry include:
  \begin{itemize}
  \item \textbf{Maxillary sinus pathology} shows very strong evidences.
  \item  \textbf{Surgery of wisdom teeth} and \textbf{intraosseous pathology} show strong evidences.
  \item \textbf{Salivary stones} can be detected very reliably.
  \item \textbf{Maxillofacial traumatology}
  \item \textbf{Assistance in image-guided surgery} \cite{scarfe2008cone}.
  \item Application of hard tissue and air cavities of the dental and ear-nose-throat areas \cite{pauwels2015cone}.
  \item Software applications in surgical simulations for \textbf{osteotomies and distraction osteogenesis} (e.g. Maxilim)
  \cite{scarfe2008cone}.
  \item \textbf{Paranasal sinus evaluation:} 
  When patients have orofacial pain symptoms, their headaches have to be evaluated which may cause paranasal pathology. CBCT is useful to image the paranasal sinus region to diagnose, e.g. pansinusitis
  \cite{miles2013atlas}.
  \item Airway analysis and diagnosis in the upper airways (e.g. in apnea).
  \end{itemize}
\end{framed}
%
\textbf{Prosthodontics} pertain the replacement of missing teeth to restore dental function \cite{Fence2013} but has no evidence for applying CBCT.

Additionally, CBCT has no application or detection in \textbf{cariology} \cite{semper2016}.

\section{Conclusion} 

This paper described the basic principles of CBCT with its applications in dental radiography. CT was introduced to receive 2D images by reconstruction 50 years ago. CBCT has indicated more benefits than image acquisition of CT, e.g. reduced scanning time and radiation dose and high image accuracy. \\

In contrary, CBCT has drawbacks in imaging soft tissue and besides that artifacts and image noise are achievable. The described technique CBCT has been shown good results in various applications in dental radiography. Especially in oral and maxillofacial imaging of hard tissue. Moreover, CBCT in dental radiography is emphasized as an indispensable tool in implantology and endondotology. Maxillofacial imaging has been developed from diagnosis to image assistance in surgeries by using specific software.\\

{\color{red}


}

\bibliographystyle{plain}	
\bibliography{chapters/bibfile}

\end{document}